%% file: main.tex
\documentclass[twocolumn]{article}
\usepackage{graphicx}
\usepackage{authblk}
\usepackage[left=2cm,right=2cm]{geometry}
\usepackage{amsmath, amscd, amssymb, amsthm}

\usepackage{xcolor}
\usepackage[normalem]{ulem} 

\newif\ifrev
\revtrue        

\usepackage{caption}
\usepackage{subcaption}
\usepackage[title]{appendix}
\usepackage{multirow}

\title{CNN-Based Locked Mode Detection and Prospects for Prediction Utilizing Density Fluctuation Signals}
\author[1]{Randall Clark} 
\author[2]{Maxim Nurgaliev} 
\author[2]{Georgy Subbotin} 
\author[3]{Jie Chen} 
\author[3]{Terry L. Rhodes} 
\author[1]{Dmitri M. Orlov}
\affil[1]{Center for Energy Research, University of California San Diego, La Jolla, CA, 92093}
\affil[2]{Next Step Fusion S.a.r.l., Bertrange, Luxembourg}
\affil[3]{University of California Los Angeles, Los Angeles, CA, 90095}

\date{}

\begin{document}

\twocolumn[
  \begin{@twocolumnfalse}
    \maketitle
    \begin{abstract}
    The deceleration of Magnetohydrodynamic (MHD) modes in tokamak plasmas is an observed behavior commonly associated with growing mode amplitude. As rotating MHD modes decelerate to a non-rotating state, they can form Locked Modes (LMs), which can degrade plasma confinement and cause disruptions. LMs therefore pose a serious threat to the sustained operation of otherwise-stable plasmas in Fusion Pilot Plants (FPPs). Identification of MHD modes and any subsequent locking is typically performed with an array of magnetic signals; these magnetic signals may not survive in the harsh reactor environment of an FPP and external magnetic measurements are delayed by magnetic diffusion through the tokamak shielding \cite{moreau2009design,biel2022development}. This work uses the Radial Interferometer Polarimeter (RIP) as an alternative diagnostic to identify the slowing of rotating MHD modes into LMs. Because this fluctuation-based approach relies on observing prior mode rotation, detection and prediction are restricted to LMs with prior rotation and exclude born-locked modes. Labels were constructed to train a Convolutional Neural Network (CNN) to distinguish MHD modes that lock from those that may appear to approach a locking state but ultimately do not lock. The trained CNN learned to identify low-frequency, descending MHD modes but struggled to distinguish modes that would continue toward locking from those that would not. A second-stage detection method significantly improved performance, detecting 88.5\% of LMs with a precision of 83.7\% by applying persistence and classification-margin filters to the CNN output; false positives can arise from low-frequency, decelerating MHD modes that trend toward locking but do not ultimately lock. These results demonstrate the capability of an FPP-relevant, non-magnetic diagnostic for accurate LM detection while also identifying fundamental limitations in the achievable prediction time using a fluctuation-based approach.
    \end{abstract}
  \end{@twocolumnfalse}
]

\input{Section1}
\input{Section2}
\input{Section3}
\input{Section4}
\input{Section5}
\input{Section6}
\input{DataAvailibility}
\input{Section7}
\input{Section8}

\bibliographystyle{plain}

\clearpage
\appendix

\begin{figure*}[!t]
    \centering

    \refstepcounter{section}
    \section*{Appendix \thesection: Dataset Coverage}

    \vspace{0.5cm}

    \includegraphics[
        width=\textwidth,
        keepaspectratio
    ]{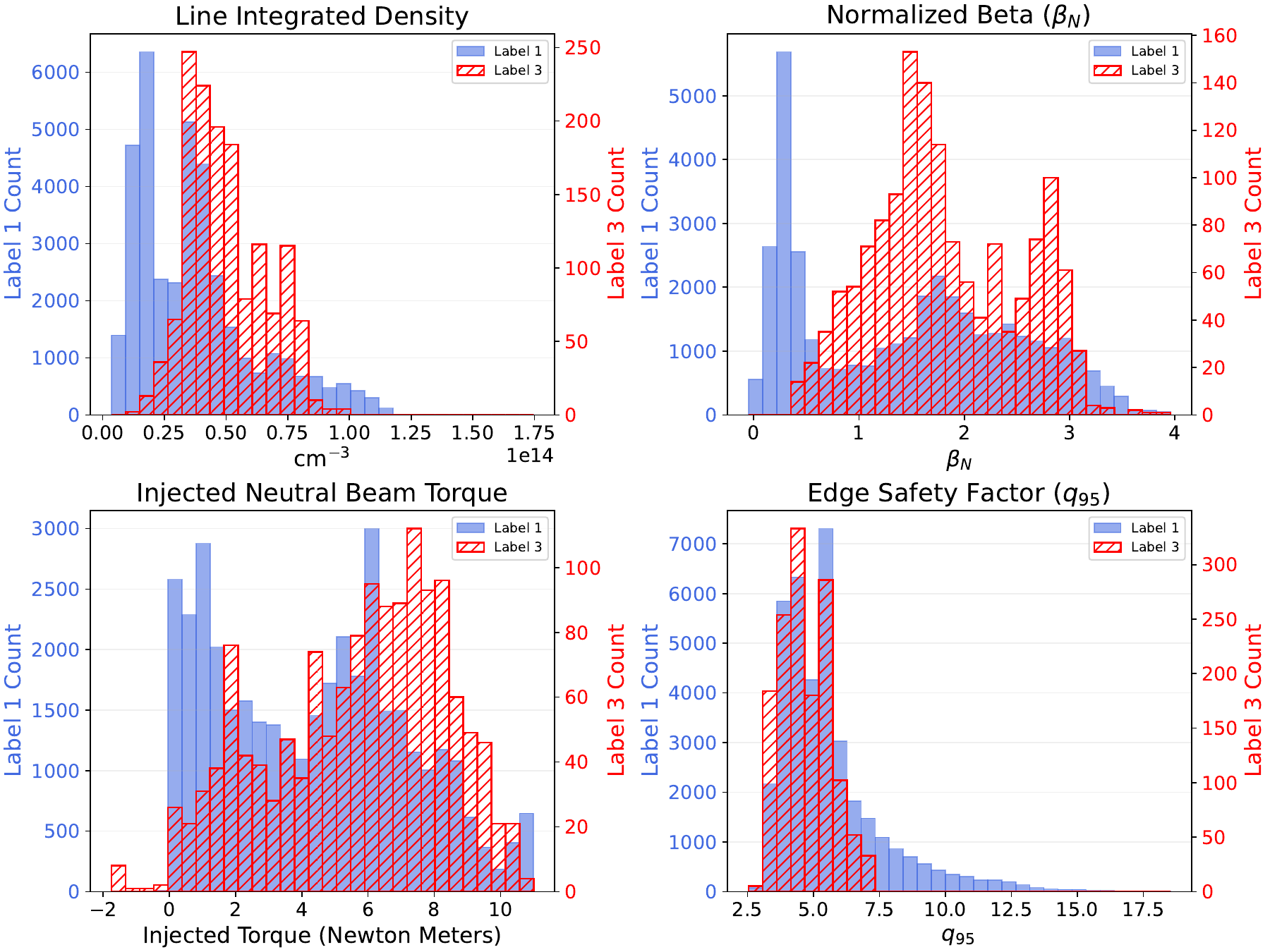}

    \caption{This collection of histograms outlines
    the dataset coverage of locked-mode-relevant
    physics parameters. Note that the injected torque
    is a time-averaged metric (averaging over the
    previous 50 ms).}
    \label{HistogramsParameters}
\end{figure*}

\end{document}

%% file: Section1.tex
\section{Introduction}
As the fusion community works toward a Fusion Pilot Plant (FPP), the need for FPP-relevant tools grows. Many of the diagnostics used by DIII-D and other tokamaks for plasma analysis and control will not be operable in an FPP due to either limited port space or a reactor environment hostile to their function. The focus of this work is Locked Mode (LM) detection and prediction using the FPP-relevant Radial Interferometer Polarimeter (RIP) \cite{chen2026multifunctional}. LM analysis is typically conducted with magnetic signals \cite{kim1999mhd,ding2014analytical}; despite their ubiquitous use in magnetic confinement fusion, these diagnostics may not survive the reactor environment and external magnetic measurements will be limited by diffusion through the tokamak shielding \cite{moreau2009design,biel2022development}. RIP offers a way to observe rotating MHD modes through density fluctuations, and through this, a way to analyze LMs with prior rotation. This study will omit the analysis of born-locked modes, LMs that emerge from the noise floor without ever showing significant rotation. Additionally, the study will be agnostic to the poloidal and toroidal mode number, only observing the presence of MHD modes and analyzing their behavior.

It has been observed across many tokamaks that MHD modes can exhibit a reduction in frequency as they grow in amplitude; when the frequency drops to zero, or near zero, the mode is considered locked. The rotation of magnetic islands is governed by the nonlinear interactions between the island, 3D magnetic fields, plasma rotation, and the surrounding conducting wall \cite{la1992critical, nave1990mode,smolyakov1995rotating}.

Locked Modes are dangerous to the operation of a tokamak as they are known to cause disruptions and degrade confinement. Sweeney et al. \cite{sweeney2017statistical} found that approximately 18\% of disruptions at DIII-D were attributable to locked modes with rotating precursors. The physics of how LMs cause disruptions is not fully understood, but it has been proposed that LMs can cause a severe collapse in confinement that precedes and may trigger the thermal quench \cite{sweeney2018relationship}. The disruptions caused by LMs pose a serious threat to FPPs and next-generation tokamaks; as a result, LM control and avoidance are significant control targets in their design \cite{maviglia2023studies, nies2022stabilisation, sweeney2020mhd}.

Various control techniques for LMs have been developed. 3D magnetic field coils have been used to compensate for the error fields that produce resonant magnetic islands \cite{scoville1991locked}. If an LM has already formed, there are techniques for unlocking it and allowing it to rotate once more to prevent disruption; this has been done with resonant magnetic perturbation techniques \cite{jin2015locked} and electrode biasing \cite{chen2021first}. LMs that are in the process of forming can be prevented through the use of electron cyclotron current drive \cite{choi2018feedforward} and magnetic perturbations \cite{volpe2015avoiding}.

Applications of machine learning (ML) to LM detection and prediction have been growing in recent years \cite{akccay2021machine,rea2026review}. Prominent examples using experimental data commonly rely on magnetic signals \cite{ferreira2022explainable,gambrioli2025machine,ding2013neural}. The work in \cite{ferreira2022explainable} demonstrates the use of CNNs to analyze magnetic spectrograms, similar to the use of CNNs to analyze RIP spectrograms in this work. Prediction of LM-induced disruptions is another active area of research \cite{deng2025interpretability}.

As LMs pose a serious threat to future tokamaks, an FPP-relevant approach to their detection and prediction is needed to enable the various LM control and unlocking techniques discussed above. This work aims to provide such a method using the FPP-relevant RIP diagnostic. Observation of mode rotation in RIP-based spectrograms has enabled CNNs to be trained to identify a mode's progression toward locking. The CNN model has been benchmarked through a 10-fold cross-validation process to strictly assess its generalizability to unseen discharges. This process divides the dataset into 10 folds, each containing chronologically contiguous discharges, with eight folds used for training, one for validation, and one for testing. Ten separate models are trained, with each fold serving as the test set exactly once. This approach enables the evaluation of model performance across different subsets of unseen discharges, providing a measure of its generalizability and variability in performance.

%% file: Section2.tex
\section{The Radial Interferometer Polarimeter}
The Radial Interferometer Polarimeter (RIP) is a high-frequency interferometer-polarimeter diagnostic capable of simultaneously measuring line-integrated electron density and magnetic-field information in the plasma. RIP uses right- and left-handed circularly polarized electromagnetic probe waves to measure the Faraday effect, allowing the polarimetric phase to be expressed as \cite{chen2021correlation}:

\begin{equation}
\varphi_{FR} = \frac{1}{2}(\varphi_{R}-\varphi_{L})
= 2.63 \times 10^{-13}\lambda^2 \int n_e B_{\parallel}dl,
\end{equation}

where $\varphi_R$ and $\varphi_L$ are the phase changes in radians of the right- and left-handed waves, respectively, $\lambda$ is the wavelength of the probe waves, $n_e$ is the electron density, and $B_{\parallel}$ is the magnetic field parallel to the probe beam. The  line integrated density is given by \cite{prabhudesai2024upgrade}:

\begin{equation}
\varphi_I = \frac{1}{2}(\varphi_R+\varphi_L)
= 2.82 \times 10^{-15}\lambda \int n_e dl.
\end{equation}

The fluctuating component of the interferometric phase therefore provides a measurement of the line-integrated electron-density fluctuation, $\delta n_e$, while the fluctuating Faraday phase provides information about magnetic-field fluctuations \cite{prabhudesai2024upgrade}.

RIP is attractive as an FPP-relevant diagnostic because it provides non-perturbing measurements of internal plasma fluctuations using electromagnetic probe beams while allowing sensitive source and detection hardware to be located remotely from the plasma. On DIII-D, RIP consists of four radial chords (although only the one chord at the midplane was used for the models constructed in this paper, making the model insensitive to the location of the mode) and has demonstrated simultaneous measurements of density and magnetic fluctuations with high temporal resolution \cite{chen2016faraday,chen2021correlation}. This provides an alternative to relying exclusively on magnetic sensing coils located outside of the plasma.

RIP was chosen for this work because of its demonstrated ability to observe MHD modes and their rotation frequencies. Previous work has used RIP to detect tearing modes and track their frequency evolution, including modes detected by RIP before they become observable using external magnetic sensing coils \cite{pandya2024early}. RIP has also been developed for toroidal mode-number measurements and the observation of high-frequency MHD activity \cite{prabhudesai2024upgrade}. Of particular relevance to this work, both magnetic and density fluctuations measured by RIP have been shown to contain coherent signatures of tearing-mode activity, including modes that decelerate and subsequently lock \cite{chen2016faraday}.

Although MHD analysis with RIP has commonly used the magnetic-fluctuation measurement obtained through Faraday-effect polarimetry, this work instead uses the interferometric density-fluctuation signal. The density-fluctuation signal, measured in units of $1/m^2$, exhibits a higher signal-to-noise ratio than the polarimetric magnetic-fluctuation signal while retaining clear signatures of the same coherent MHD activity \cite{chen2021correlation}. Additionally, the interferometric measurement avoids the processing required to infer magnetic fluctuations from the polarimetric measurement. The resulting density-fluctuation spectrograms provide the CNN with the presence and frequency evolution of rotating MHD modes, allowing the model to identify their progression toward locking.

%% file: Section3.tex
\section{Training a CNN on Spectrogram Data}
\subsection{Label Criteria For Manual Labeling}
The construction of an ML model that can detect and predict LMs required careful consideration regarding the structure of the training labels. Locking events were broken down into time periods to label the degree of danger an MHD mode presents (if one is present). Inspiration for this idea came from \cite{gambrioli2025machine}, where an unstable period of 60 ms was designated before every LM. This idea has been expanded upon to create a three-label system that follows these criteria:

\begin{itemize}
    \item Label 1 (No Warning): All times when an early warning or emergency warning is not warranted.
    \item Label 2 (Early Warning): The MHD mode is greater than 5 kHz and shows a clear, sustained decrease in rotation frequency. Label 2 may persist below 5 kHz, but cannot originate there.
    \item Label 3 (Emergency Warning): The MHD mode is less than or equal to 5 kHz, decreasing in rotation frequency, and will lock within 100 ms.
\end{itemize}

These criteria were carefully chosen so that an emergency warning is assigned only when a decreasing MHD mode is sufficiently slow and is known from the labeled data to lock within 100 ms. This choice trains the model to distinguish between low-frequency MHD modes ($<5$ kHz) that continue toward locking and those that slow but do not lock within 100 ms. This creates an additional, intended challenge for the ML model: a slowly rotating MHD mode ($<5$ kHz) that decreases in frequency but does not lock within 100 ms is assigned Label 1. This label assignment is intended to train the model to use all available information to recognize, if possible, the cues that distinguish near-miss slowing modes from those that will lock. As discussed in Section \ref{SectionFilter}, the model ultimately struggles with this distinction, suggesting that the RIP spectrogram alone may contain limited information for reliably predicting whether the mode will continue slowing.

The dataset consists of 121 shots containing 139 locked modes with prior rotation. The dataset spans approximately one year, from May 2024 through May 2025, and is limited to cases in which RIP was recording and LMs formed with prior rotation. Figure \ref{HistogramsParameters} in the appendix outlines the spread of relevant physics parameters across the dataset.

Labels were generated through manual analysis of magnetic diagnostic signals using MODESPEC and SLCONTOUR. These tools were used to identify rotating modes that decelerated and subsequently locked, according to the criteria outlined above \cite{strait2006magnetic}.

Substantial class imbalance must be accounted for in this ML problem, as the No Warning class (Label 1) comprises over 90\% of the dataset, while the most important class, Emergency Warning (Label 3), comprises only 3.5\%. To account for class imbalance during model training, a weighted cross-entropy loss function was used, with weights determined by the frequency of each class in the training set. The class weight is defined as:

\begin{equation}
w_c = \frac{N}{KN_c}
\end{equation}

where $N$ is the total number of data points, $K$ is the number of classes, and $N_c$ is the number of data points in class $c$. The weighting of the loss function ensures that each class has an equal aggregate weight during training. This helps prevent the model from becoming biased toward the most ubiquitous class.

\begin{table}[!ht]
\begin{center}
\begin{tabular}{||c c c c||} 
 \hline
 Data Imbalance & Label 1 & Label 2 & Label 3 \\ [0.5ex] 
 \hline\hline
 Total Data  &  38858 & 2302 & 1487\\ 
 \hline
 Percent Data  & 91.1\% & 5.4\% & 3.5\% \\ 
 \hline
 Class Weight  & 0.37 & 6.18 & 9.56 \\ 
 \hline
\end{tabular}
\end{center}
\caption{The dataset consists of 139 LMs across 121 shots. This table shows the breakdown of data for each of the three labels defined above. The global class weights are included for illustrative purposes; during training, these weights are recalculated using the available training data.}
\label{ClassBalanceLabels}
\end{table}

\subsection{10-fold Cross-Validation Training CNNs}
Model training was carefully set up to provide insight into how the model will perform on unseen DIII-D discharges. A common approach in ML is to randomly shuffle data into an 80-20 train-test split; however, this approach would greatly inflate test performance and give undue confidence in the model's performance on new experiments. This performance inflation would occur because DIII-D data are highly correlated across nearby time steps, resulting in significant temporal leakage between the training and test datasets. Additionally, data collected from shots on the same experimental day can contain similar features, introducing another potential source of leakage. This leakage problem is addressed with a 10-fold cross-validation setup \cite{hastie2009elements}.

The 10-fold cross-validation procedure partitions all shots into 10 folds organized chronologically. For each iteration, eight folds are used for training, one for validation, and one for testing, resulting in an 80-10-10 training-validation-test split. This ordering limits leakage caused by chronologically nearby shots with similar features appearing across the training, validation, and test sets. The validation fold is always two folds ahead of the test fold, with one training fold between them to provide additional chronological separation. Throughout this work, model selection and tuning are performed using the validation fold, while final performance is evaluated on the held-out test fold. Because no data from the test shots are included during model training or selection, these shots serve as proxies for unseen DIII-D discharges. This procedure requires training 10 separate CNNs, with aggregate test performance obtained by evaluating each model on its corresponding held-out test fold. This setup provides insight into how the trained models generalize to unseen data and new discharges.

\begin{figure}[!ht]
\centering
\includegraphics[width=8.5cm,height=8.5cm,keepaspectratio]{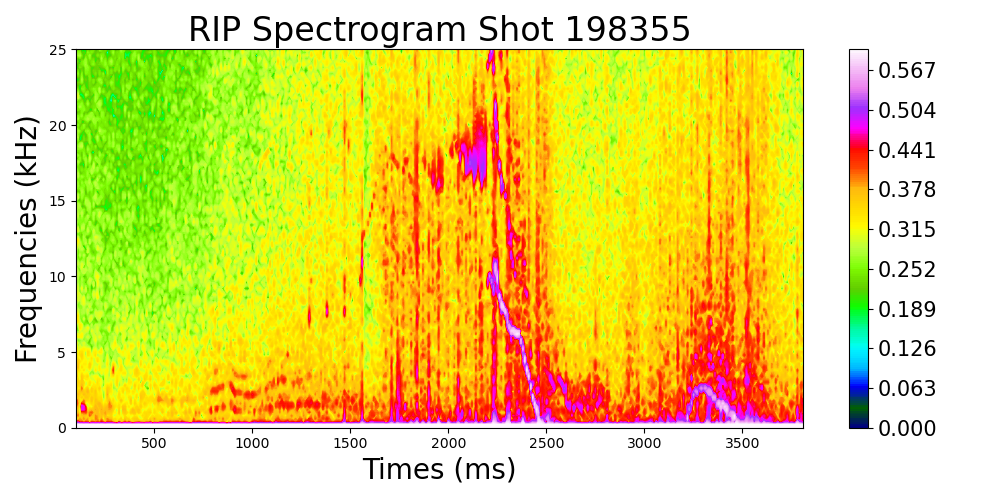}
\caption{RIP provides rapidly sampled line-integrated density measurements that can be transformed into spectral data capable of observing rotating MHD modes. Spectrogram values are calculated from the logarithm of the magnitude of the complex STFT coefficients and normalized using min-max scaling. The CNN receives 10 time slices spanning 90 ms, as shown here, for shots in the training, validation, and test datasets.}
\label{Original_Data}
\end{figure}

The data are preprocessed using min-max scaling to normalize the spectral amplitudes between 0 and 1 using global min-max parameters. A Short-Time Fourier Transform (STFT) is applied to the RIP data using 5 ms windows to produce spectra at 10 ms intervals \cite{durak2003short}. At each time step, the STFT window contains only the preceding 5 ms of data, preventing future information from leaking into the model input. The CNN input is constructed by taking the logarithm of the magnitude of the complex STFT coefficients before applying min-max normalization. Each CNN input contains 10 time slices separated by 10 ms, spanning the current time and the preceding 90 ms. The resulting CNN input has dimensions $10\times126$ (time frames $\times$ frequency bins), spanning 90 ms in time and 0–25 kHz in frequency with approximately 0.2 kHz frequency spacing. Higher frequencies available from RIP are truncated because they are not relevant to the MHD behavior considered in this work and would unnecessarily increase the model input size. See Figure \ref{Original_Data} for a depiction of the spectrogram that the CNN receives.

The 10-fold cross-validation procedure trains 10 separate CNNs, with model selection based on maximizing the validation macro F1 score. The macro F1 score is the unweighted average of the individual F1 scores across all three classes, giving each class equal importance regardless of class imbalance. The Adam optimizer and weighted cross-entropy loss function are used for training \cite{kingma2014adam}. Training is limited to a maximum of 150 epochs, with early stopping after 25 epochs without improvement in validation macro F1 to reduce overfitting and improve generalization to the test data. The model checkpoint with the highest validation macro F1 is retained for test evaluation.

\begin{figure}[!htp]
\centering
\includegraphics[width=8.5cm,height=16cm,keepaspectratio]{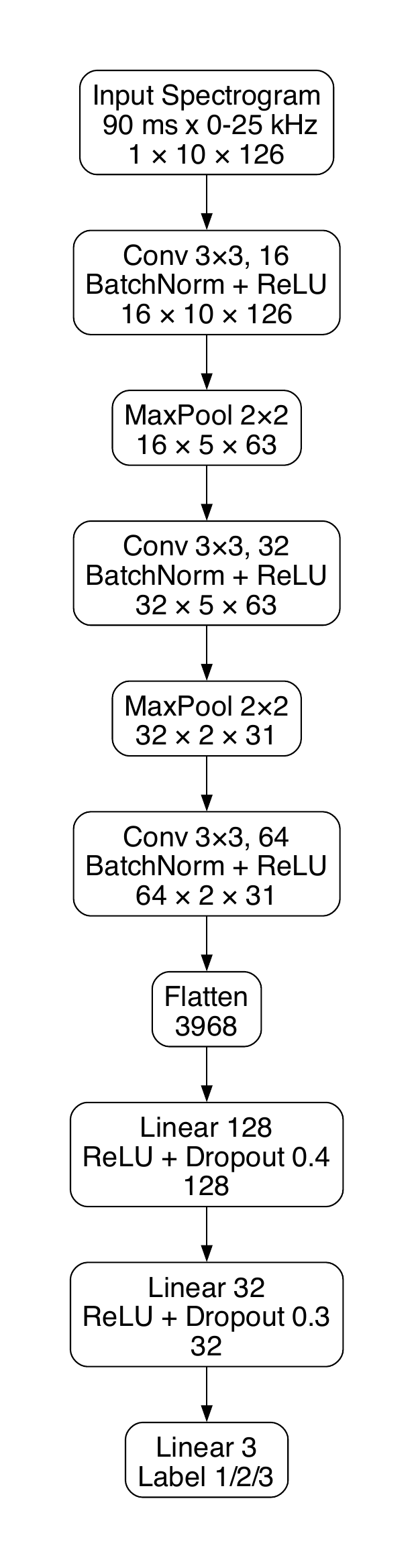}
\caption{This is the final CNN architecture selected through an ablation study that maximized validation macro F1 while varying network size and depth, pooling configuration, and input window length. This architecture provided the best validation performance while maintaining minimal model complexity.}
\label{CNN_Model}
\end{figure}

The CNN architecture follows a standard structure using $3\times3$ convolutional kernels and alternating convolutional and pooling layers \cite{simonyan2014very}. An ablation study was performed to select the best-performing architecture on the validation data while minimizing model complexity for computational efficiency. The various approaches in the ablation study attempted to capture the curves of descending MHD modes; pooling size, network depth and size, and input window length were varied. Additional preprocessing methods were also considered. Denoising using median filtering and ridge identification using Frangi filters were evaluated but failed to improve performance while increasing computational overhead \cite{frangi1998multiscale}. The final model is displayed in Figure \ref{CNN_Model}.

After training all 10 CNNs for the 10-fold cross-validation experiment, the results can be seen in Table \ref{CNN_Performance}. Parameters of the preprocessing and training are shown in Table \ref{Model_Parameters}.

\begin{table}[!ht]
\begin{center}
\begin{tabular}{||c c c||} 
 \hline
 CNN Performance & Validation & Test \\ [0.5ex] 
 \hline\hline
 Accuracy  & $94.5 \pm 3.1\%$  & $93.6 \pm 3.4\%$ \\ 
 \hline
 Balanced Accuracy  & $82.0 \pm 4.6\%$  &  $79.7 \pm 3.7\%$\\ 
 \hline
 Macro Precision  & $77.5 \pm 5.4\%$  &  $75.1 \pm 7.6\%$\\ 
 \hline
 Macro Recall  & $82.0 \pm 4.6\%$  &  $79.7 \pm 3.7\%$\\ 
 \hline
 Macro F1  & $79.4 \pm 4.8\%$  &  $76.5 \pm 5.1\%$\\ 
 \hline
 Label 3 Precision  & $76.3 \pm 11.0\%$  &  $73.1 \pm 14.2\%$\\ 
 \hline
 Label 3 Recall  & $82.0 \pm 6.4\%$  &  $77.3 \pm 6.9\%$\\ 
 \hline
 Label 3 F1  & $78.6\ \pm 7.4\%$  &  $74.1 \pm 7.0\%$\\ 
 \hline
\end{tabular}
\end{center}
\caption{CNN performance for the validation and held-out test folds after model selection based on validation macro F1. Macro-averaged metrics are reported along with Label 3 metrics because Label 3 is the primary class of interest for LM detection and prediction. Values represent the mean and standard deviation across the 10 cross-validation folds.}
\label{CNN_Performance}
\end{table}

\begin{table}[!ht]
\begin{center}
\begin{tabular}{||c c||} 
 \hline
 Model Parameters & Value \\ [0.5ex] 
 \hline\hline
 MIN  & 0\\ 
 \hline
 MAX  & 8.96\\ 
 \hline
 Window length  & 90 ms\\ 
 \hline
 Frequency Range  & 0--25 kHz\\ 
 \hline
 CNN input dimension  & $10\times126$\\ 
 \hline
 Optimizer  & Adam \\ 
 \hline
 Loss Function  & Weighted Cross Entropy\\ 
 \hline
 Learning Rate  & $10^{-3}$\\ 
 \hline
 Batch Size  & 128\\ 
 \hline
 Max Epochs  & 150\\ 
 \hline
 Patience  & 25\\ 
 \hline
\end{tabular}
\end{center}
\caption{Min-max normalization is applied to the log-magnitude spectral data using global normalization bounds of 0 and 8.96. Values below the lower bound are clipped to exclude excessively small spectral amplitudes; no upper-bound clipping is applied. The remaining parameters describe the CNN input and neural-network training procedure.}
\label{Model_Parameters}
\end{table}

\subsection{CNN Performance and Errors}
The performance of the CNN on the test data outlined in Table \ref{CNN_Performance} indicates that the model struggles to consistently align with the pre-defined labels. An exploratory analysis was conducted to identify possible causes of this non-ideal performance, offering both statistical and anecdotal insight into the CNN's behavior. This analysis also provides guidance for improving LM detection and prediction.

\begin{figure}[!ht]
\centering
\includegraphics[width=8.5cm,height=8.5cm,keepaspectratio]{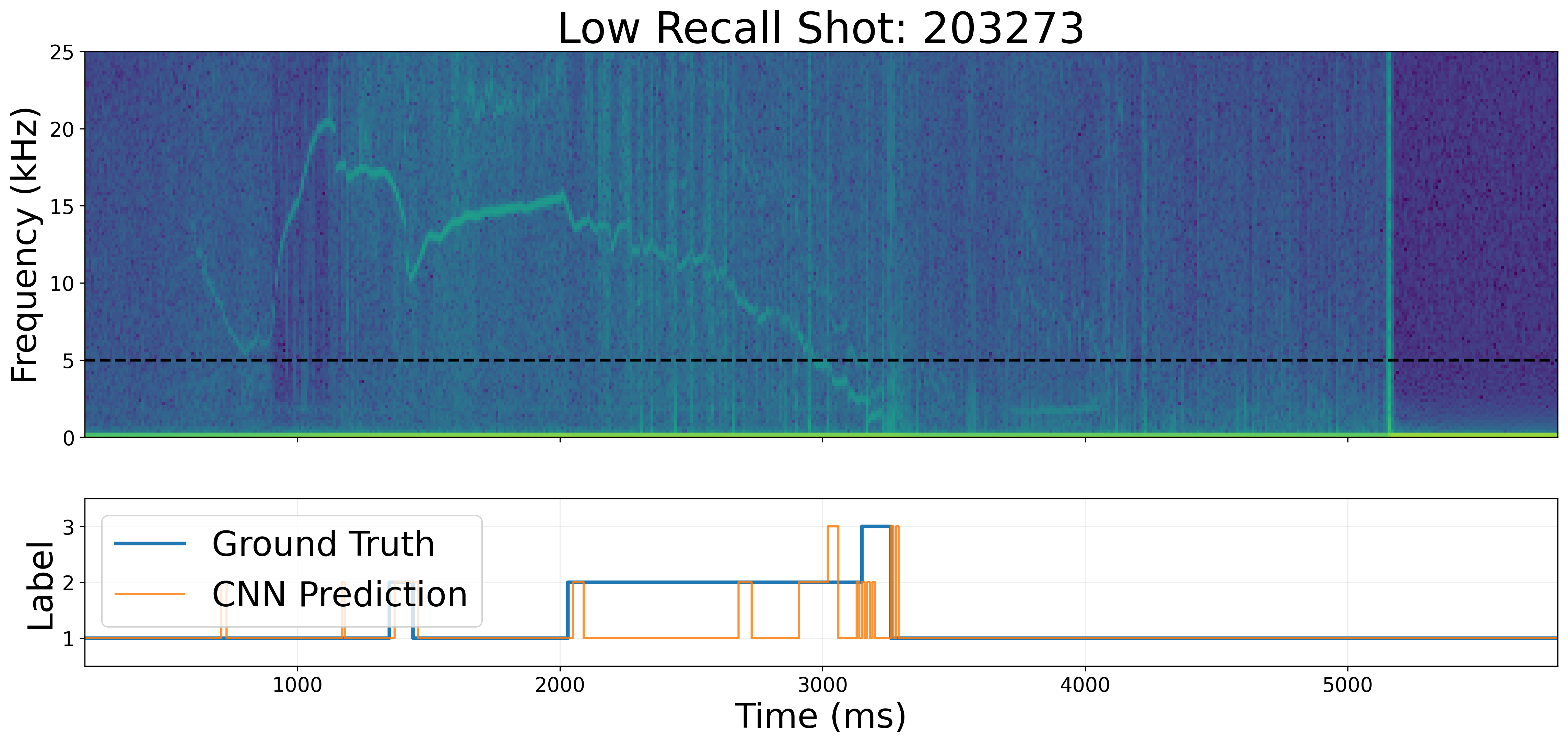}
\caption{This is an example of a low-recall shot. The upper plot shows the RIP spectrogram and the lower plot lists the class output of the CNN in the test fold. This plot demonstrates a limitation of using global classification statistics for an event-detection problem: the CNN is alert to the LM both before and after the true Label 3 interval despite failing to consistently align with the predefined labels. This example is particularly difficult for the CNN because the MHD mode exhibits step-like frequency evolution; without a clear negative slope, the CNN struggles to consistently identify the imminent LM. While the predictions are temporally misaligned with the true labels, nearby Label 3 predictions can still be used to detect the LM, as discussed in Section \ref{SectionFilter}.}
\label{LowRecallPlot}
\end{figure}

\begin{figure}[!ht]
\centering
\includegraphics[width=8.5cm,height=8.5cm,keepaspectratio]{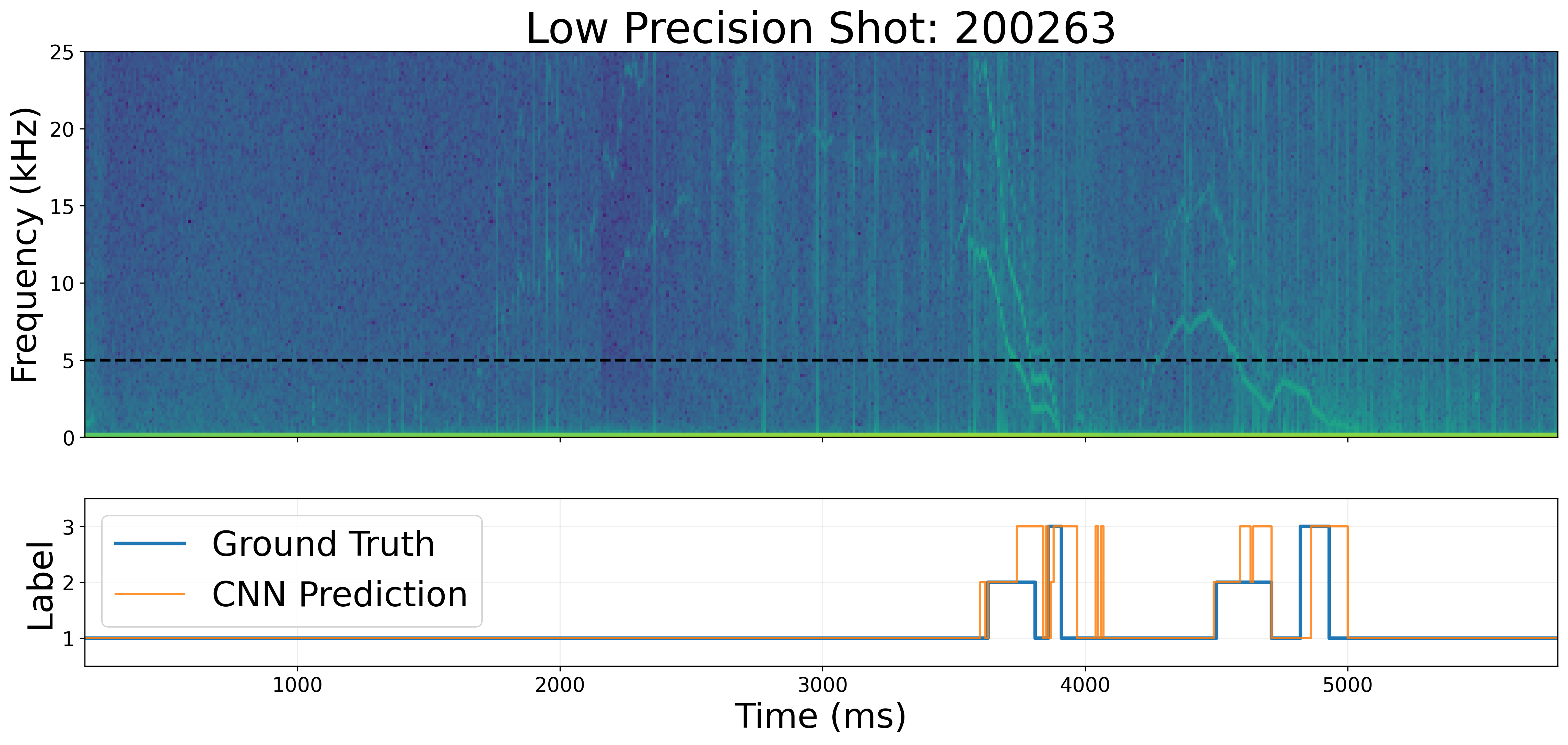}
\caption{This is an example of a low-precision shot. The upper plot shows the RIP spectrogram, while the lower plot shows the class output of the CNN in the test fold. While this is only one example, it provides insight into how the CNN assigns its labels. At approximately 3800 ms, the CNN raises an emergency warning (Label 3), without knowing that the MHD mode frequency will briefly remain steady before continuing its descent into an LM at approximately 4000 ms; similar behavior occurs again at 4600 ms. This indicates that the CNN can identify low-frequency MHD modes that are rapidly descending but cannot reliably predict whether that behavior will continue. Erroneous single-time-slice Label 3 predictions around 4100 ms result from noisy perturbative data influencing the CNN output; Section \ref{SectionFilter} addresses how these predictions are handled.}
\label{LowPrecisionShot}
\end{figure}

An error analysis was conducted by examining shots with the lowest recall (fraction of actual positives correctly identified) and lowest precision (fraction of positive predictions that are correct). Examples of a low-recall and a low-precision shot are shown in Figures \ref{LowRecallPlot} and \ref{LowPrecisionShot}, respectively. While anecdotal, these examples provide insight into how the model interprets the data, indicating that it is sensitive to low-frequency descending modes irrespective of their future evolution. Section \ref{SectionFilter} uses the behavior observed in these examples to guide the construction of a more accurate LM detection method built around the CNN output.

\begin{figure}[!ht]
\centering
\includegraphics[width=8.5cm,height=8.5cm,keepaspectratio]{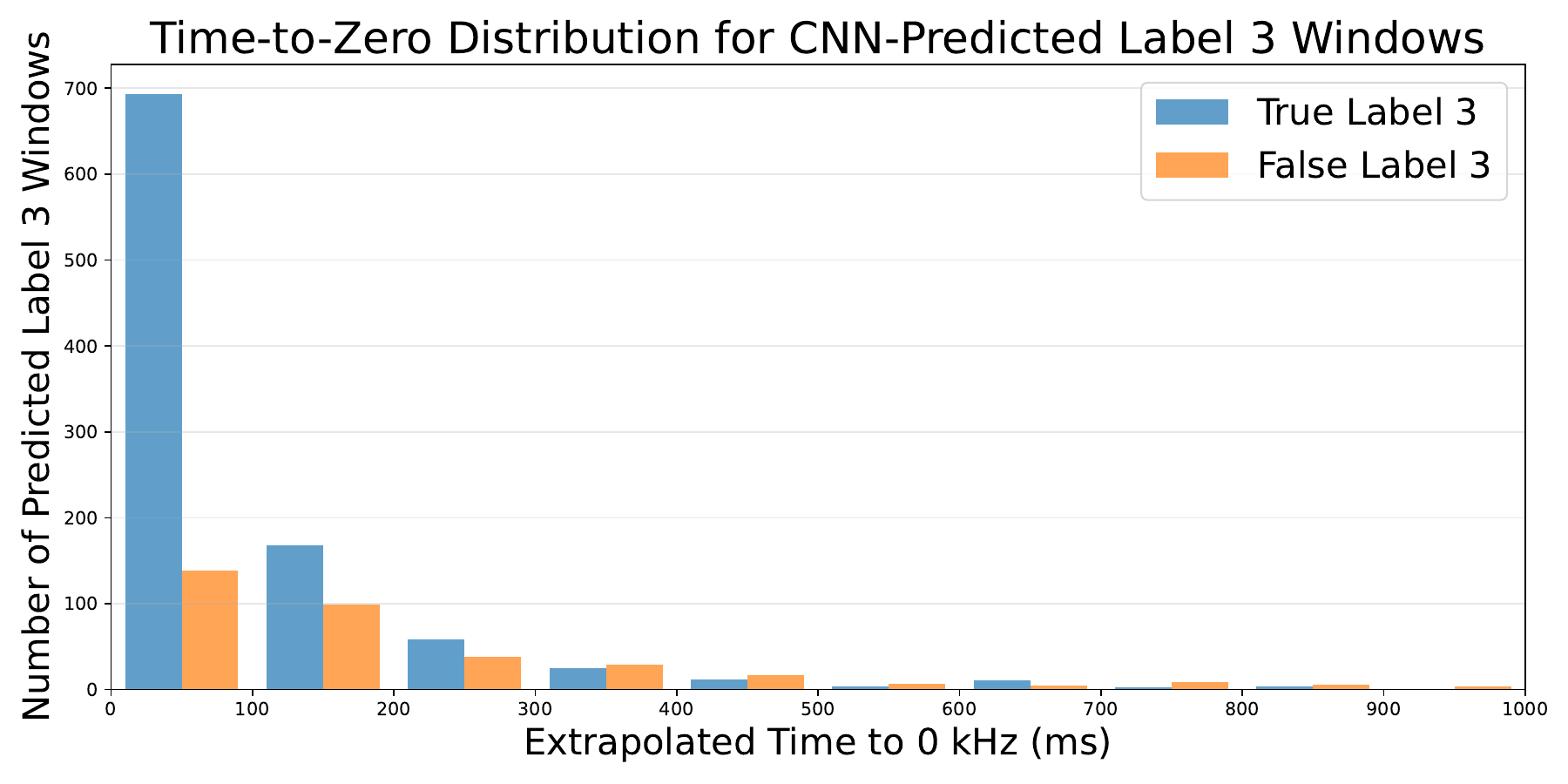}
\caption{The linear slope extrapolation analysis compares the time of the last time slice in the predicted Label 3 window to when the fitted line intersects with the x-axis where frequency would be 0 and an LM would be predicted to form. Even when the CNN incorrectly predicts a Label 3, it does so because it frequently sees a descending MHD mode. As shown in Table \ref{Table_SlopeTo0Analysis}, 85\% of true-positive predictions and 64\% of false-positive predictions linearly extrapolate to 0 kHz within 1000 ms. The limitations of this crude linear extrapolation should be emphasized; the analysis is intended to characterize the frequency evolution associated with CNN predictions rather than provide an accurate estimate of the physical locking time.}
\label{SlopeTo0Analysis}
\end{figure}

A physically relevant feature available to the CNN is the downward frequency evolution associated with the slowing of an MHD mode toward locking. A crude representation of this behavior can be obtained by fitting a line to the lowest-frequency spectral ridge within each window predicted as Label 3 in the test set. A simple peak-finding algorithm identifies spectral peaks at each time slice, from which the lowest-frequency peak is selected to represent the mode. These points are then fit with a line. For windows with a negative fitted slope, extrapolating the line to 0 kHz provides a rough estimate of the time at which the mode would lock if the observed frequency evolution continued linearly. Recall that the Label 3 training criteria require the LM to form within $\leq100$ ms. Figure \ref{SlopeTo0Analysis} shows the distribution of extrapolated times to 0 kHz, illustrating the tendency for both true-positive and false-positive Label 3 predictions to contain rapidly descending MHD modes.

Table \ref{Table_SlopeTo0Analysis} lists the results of this analysis and shows that both true-positive and false-positive Label 3 predictions have a strong bias toward windows containing a decreasing MHD mode. As expected, the true-positive predictions show a stronger bias toward descent. Positive fitted slopes can result from noisy window data, blunt thresholding, or, in the case of false-positive predictions, MHD modes rebounding to higher frequencies after previously descending. This analysis is intentionally crude and its results are not definitive, but it provides additional insight into the behavior learned by the CNN.

\begin{table}[!ht]
\begin{center}
\begin{tabular}{||c c||} 
 \hline
 Extrapolated Time to 0 kHz & Value \\ [0.5ex] 
 \hline\hline
 True L3, $\leq$ 1000 ms  & 85\%\\ 
 \hline
 True L3, $\leq$ 300 ms  & 79\%\\ 
 \hline
 False L3, $\leq$ 1000 ms  & 64\%\\ 
 \hline
 False L3, $\leq$ 300 ms  & 50\%\\ 
 \hline
 Threshold  & 0.5\\ 
 \hline
 Skip  & 4\\ 
 \hline
\end{tabular}
\end{center}
\caption{The results and method parameters for the linear extrapolation analysis are recorded here. Lock times listed in this table are estimations based on linear extrapolation of fitted lines, not true locking times. The results are for the true-positive and false-positive Label 3 predictions by the 10 CNNs on their respective test data. The Threshold parameter specifies the minimum normalized spectral amplitude required for inclusion in the $(X,Y)$ fit. Skip excludes the first four spectrogram entries, which generally contain large low-frequency components that do not correspond to the MHD modes of interest.}
\label{Table_SlopeTo0Analysis}
\end{table}

\begin{figure}[!ht]
\centering
\includegraphics[width=8.5cm,height=8.5cm,keepaspectratio]{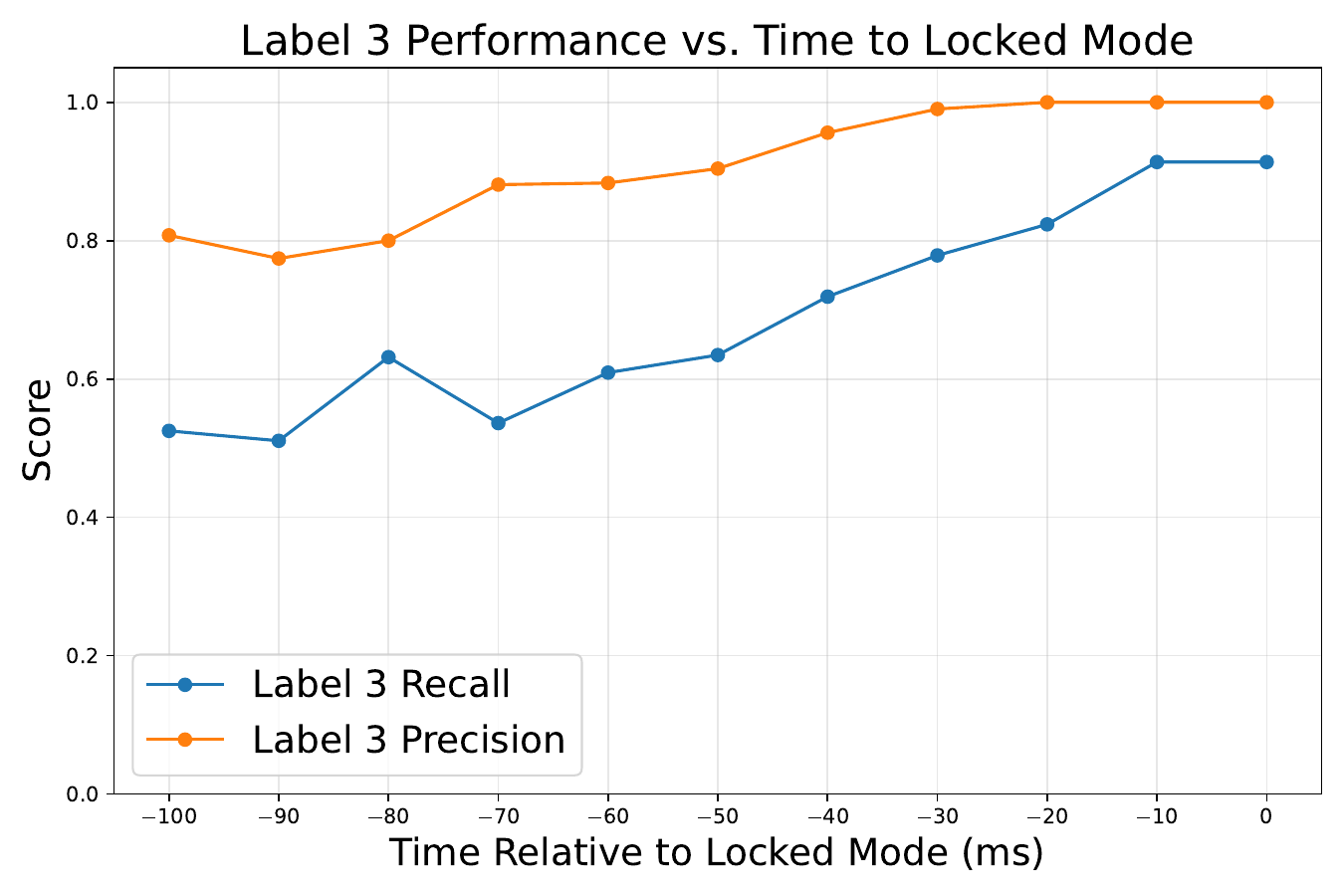}
\caption{CNN precision and recall improve as MHD modes approach the locking state. The recall provides the more informative trend: as a mode approaches locking, the CNN becomes increasingly likely to identify it as an emergency warning. Statistics are aggregated across the 10 CNNs evaluated on their respective held-out test folds.}
\label{CNN_Time_To_Lock_Analysis}
\end{figure}

As the emergency warning label is the primary focus of this tool, it is informative to examine its performance as a function of time to locking. Figure \ref{CNN_Time_To_Lock_Analysis} shows the precision and recall of the CNN throughout the final 100 ms preceding an LM. Precision reaches 100\% immediately before the LM, which has to happen by definition (as all LMs in the dataset have a Label 3 immediately before them, negating the possibility of a false positive at this time). Recall increases as the locking event approaches, reaching 91.4\% immediately before locking.

%% file: Section4.tex
\section{CNN Filters and Locked Mode Detection} \label{SectionFilter}
\subsection{Identifying Filters for Improved Locked Mode Detection}
The CNN analysis performed in the previous section indicates that the locked mode detector performs better than the global statistics listed in Table \ref{CNN_Performance} would imply. Inspired by the prior analysis, an approach based on event detection, rather than time-slice accuracy, is developed. This approach is defined by a second-stage detection method referred to as the "Block Approach," which examines collections of contiguous predicted Label 3s rather than individual time-slice predictions. If there exists even a single Label 1 or Label 2 prediction between sets of Label 3 predictions, the two blocks of Label 3s are treated as separate blocks. This approach evaluates the metrics associated with blocks of Label 3s rather than the individual metrics associated with the time slices that comprise those Label 3 predictions.

Blocks of Label 3s are considered true-positive blocks if they contain at least one correctly predicted Label 3 time slice. This approach was found to be the most effective, but creates two issues. The first issue is that if the CNN predicts a continuous sequence of Label 3s containing at least one true Label 3, the entire sequence will be considered a true-positive block despite potentially persisting for a nonphysical length of time. Analysis in Section \ref{SectionPredictionCapability} shows this not to be an issue, and an upper limit on block duration could be imposed if necessary. The second issue arises when two or more nearby true-positive blocks correspond to the same LM. In this work, these blocks are not counted as multiple true-positives; instead, blocks corresponding to the same LM are counted as a single successful detection, including in the results presented in Table \ref{BlockDetection}. In practice, closely spaced blocks could provide an additional source of error that could be mitigated by limiting how close separate LM detection times can occur. Table \ref{TimeofBlockDetection} demonstrates the accuracy with which LM times can be calculated to support such an approach.

\begin{figure}[!ht]
\centering
\includegraphics[width=8.5cm,height=8.5cm,keepaspectratio]{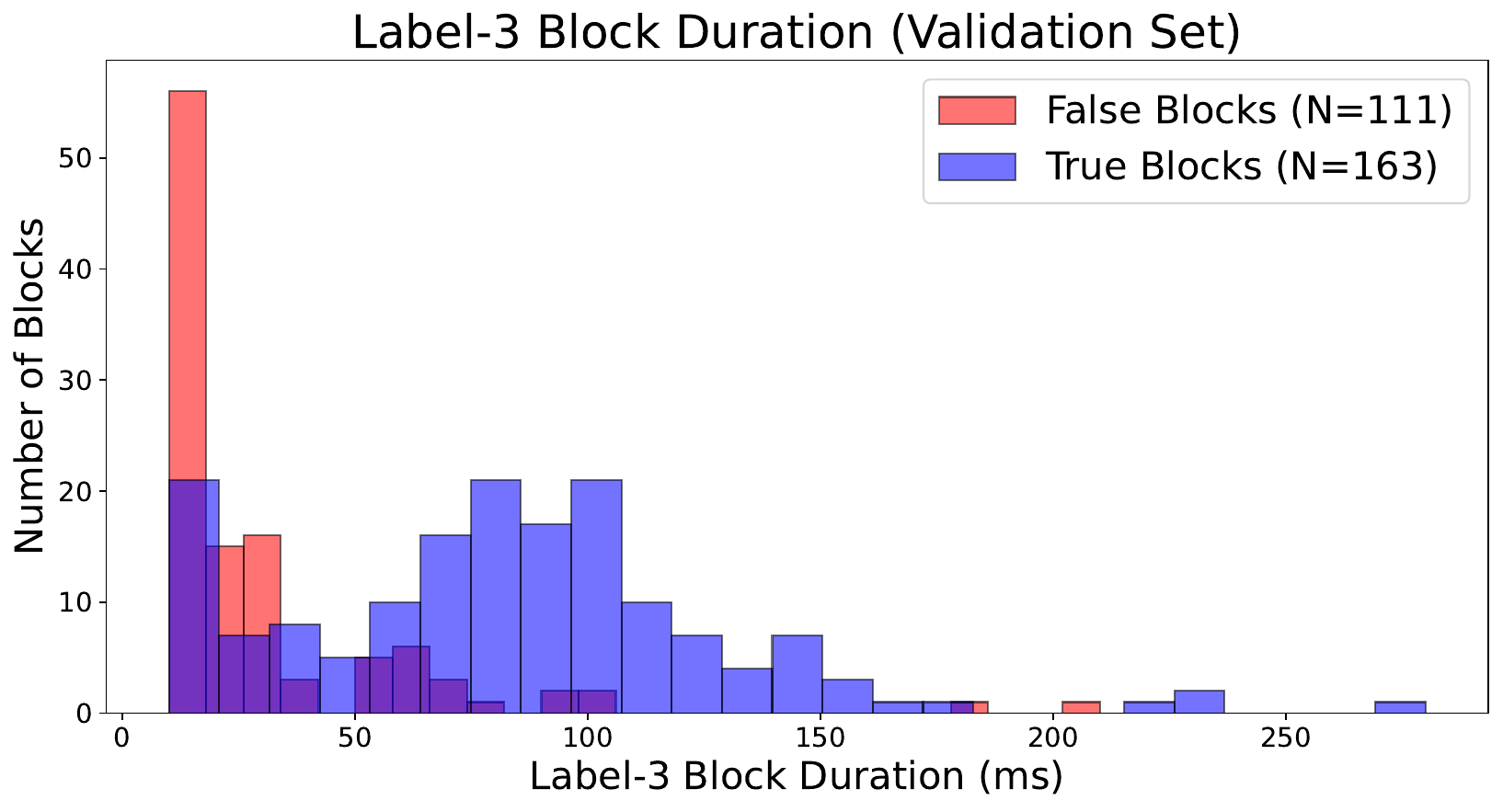}
\caption{This histogram depicts the significant number of short-duration false-positive blocks and the high concentration of long-duration ($\geq50$ ms) true-positive blocks. The clear separation between the blocks provides strong support for duration as a filtering metric for improving LM detection.}
\label{Histogram_Block_Duration}
\end{figure}

\begin{figure}[!ht]
\centering
\includegraphics[width=8.5cm,height=8.5cm,keepaspectratio]{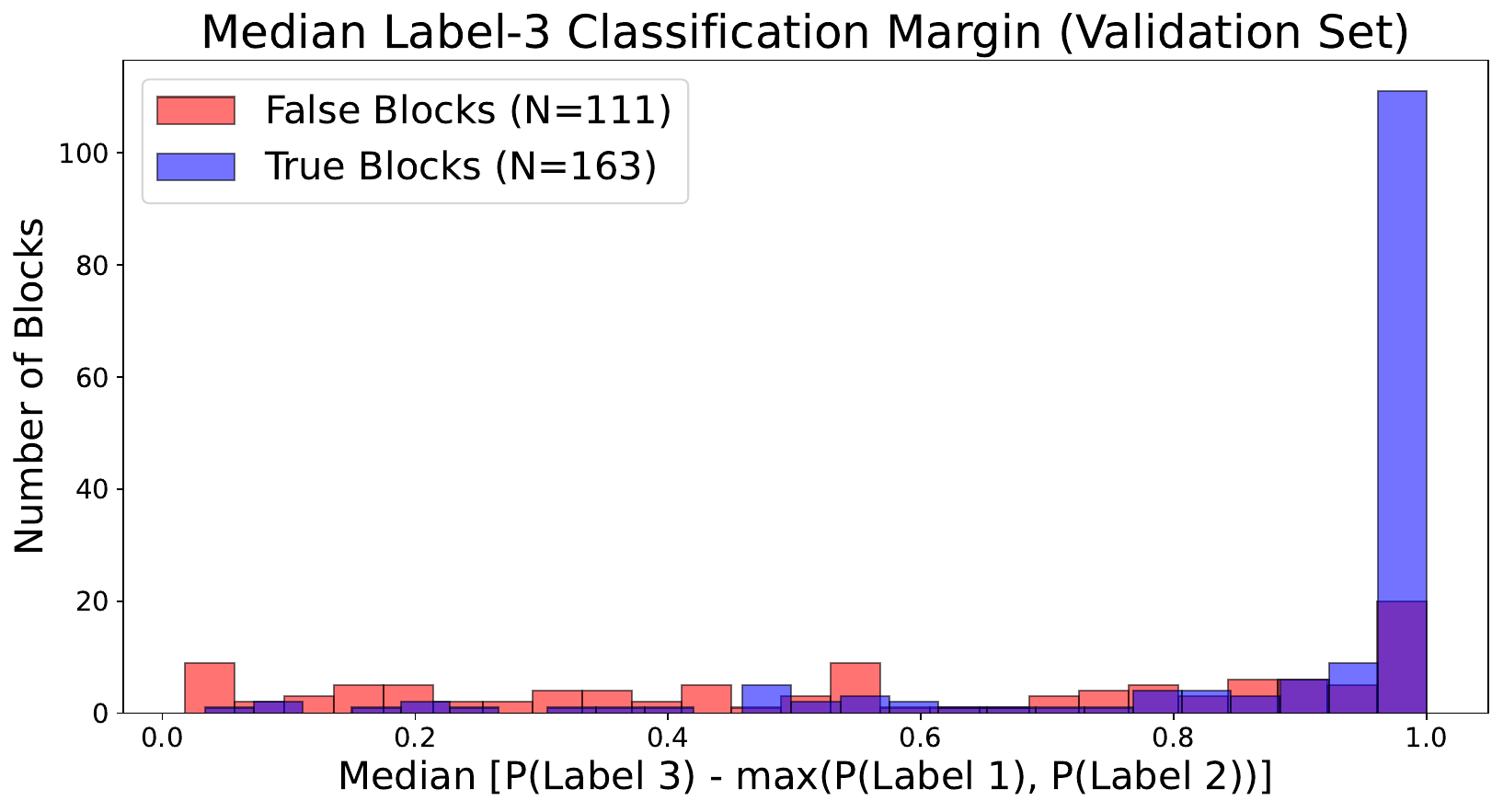}
\caption{This histogram displays the difference in median margin between the true and false-positive blocks. It is evident that the true-positive blocks are associated with a stronger probability metric, and this can be used to filter out some of the false-positive blocks, but not as acutely as the duration filter.}
\label{Histogram_Block_Margin}
\end{figure}

Collecting all contiguous blocks of Label 3 predictions results in 274 blocks in the validation set and 335 blocks in the test set. Feature selection and filter development were performed exclusively using the validation blocks. A broad analysis of statistical and model-based features was conducted on the validation blocks to identify features capable of separating false-positive from true-positive Label 3 blocks. This analysis revealed two prominent features for filtering false-positive blocks. The first and more significant feature is block duration, defined as the length of time for which the Label 3 prediction persists. A significant number of false-positive blocks persist for only 10 or 20 ms and are assumed to result from large perturbations or noise in the data. The second, less significant feature is the median classification margin across the block. At each time slice, the margin is calculated as the difference between the predicted probability of Label 3 and the larger of the predicted probabilities for Labels 1 and 2. These results are outlined in Figure \ref{Histogram_Block_Duration} and Figure \ref{Histogram_Block_Margin}.

A coarse parameter sweep was performed across the validation predictions to identify the filter parameters; recall that the 10-fold cross-validation procedure consists of 10 separately trained CNNs, each with a corresponding validation fold. These same folds and CNNs were retained for this analysis. To choose the persistence threshold, a region of high validation F1 score (88--90\%) was identified for persistence values of 40--60 ms. A value of 50 ms was selected as the midpoint of this high-performance region. A similar approach was used for the classification margin, with an optimal plateau identified between 0.2 and 0.4. A margin threshold of 0.3 was therefore selected. With these filters applied, the final LM detector identifies 88.5\% of LMs with a precision of 83.7\%, representing a substantial improvement over the CNN-only approach presented in the previous section.

\begin{table}[!ht]
\begin{center}
\begin{tabular}{||c c c||} 
 \hline
 Block Detection & Validation & Test \\ [0.5ex] 
 \hline\hline
  Precision &  87.4\% & 83.7\%\\ 
 \hline
 Recall &  89.9\% & 88.5\%\\ 
 \hline
 F1 &  88.7\% & 86.0\%\\ 
 \hline
 Blocks before filtering & 256 & 310\\ 
 \hline
 Blocks after filtering &  143 & 147\\ 
 \hline
 Persistence & 50 ms  & 50 ms\\ 
 \hline
 Margin &  0.3 & 0.3\\ 
 \hline
\end{tabular}
\end{center}
\caption{The results presented in this table show the performance of the final block-based LM detector. Metrics are calculated by aggregating all 139 LM events across all folds rather than averaging fold-level statistics; therefore, standard deviations across folds are not reported. The filter values used for block acceptance, tuned exclusively on the validation set, are also presented.}
\label{BlockDetection}
\end{table}

\subsection{Estimating Lock Times}
Determining the locking time of an LM given a true-positive block is a straightforward process that can be accurately achieved with a simple algorithm. Using the find\_peaks algorithm from SciPy \cite{2020SciPyNMeth}, the MHD modes can be identified by the prominence of their peaks. These peaks can be followed during their descent into the formation of an LM. The algorithm goes as follows:

\begin{itemize}
    \item Start at the first time slice, $t_0$, of the true-positive block and apply the find\_peaks algorithm to the normalized RIP spectrum at this time.
    \item Record the lowest-frequency peak as $f_0$. If no peak is found, iterate forward in time until one is identified.
    \item Move forward one 10 ms time step and identify the spectral peaks. If any peaks exist such that $f_{\mathrm{new}}\leq f_0+2$ kHz, record the lowest peak frequency as the new $f_0$ and continue tracking forward in time.
    \item If no peak satisfies this condition, retain the previous $f_0$ and increment a consecutive-failure counter. Continue to the next 10 ms time step.
    \item If a subsequent time step satisfies the tracking condition, update $f_0$ and reset the consecutive-failure counter to zero.
    \item When the consecutive-failure counter reaches a specified limit, terminate the ridge tracking and return the time of the last successful detection plus a flat shifting parameter as the estimated LM time.
    \item Ridge tracking is restricted to a maximum search window extending 100 ms beyond the end of the predicted Label 3 block.
\end{itemize}

\begin{figure}[!ht]
\centering
\includegraphics[width=8.5cm,height=8.5cm,keepaspectratio]{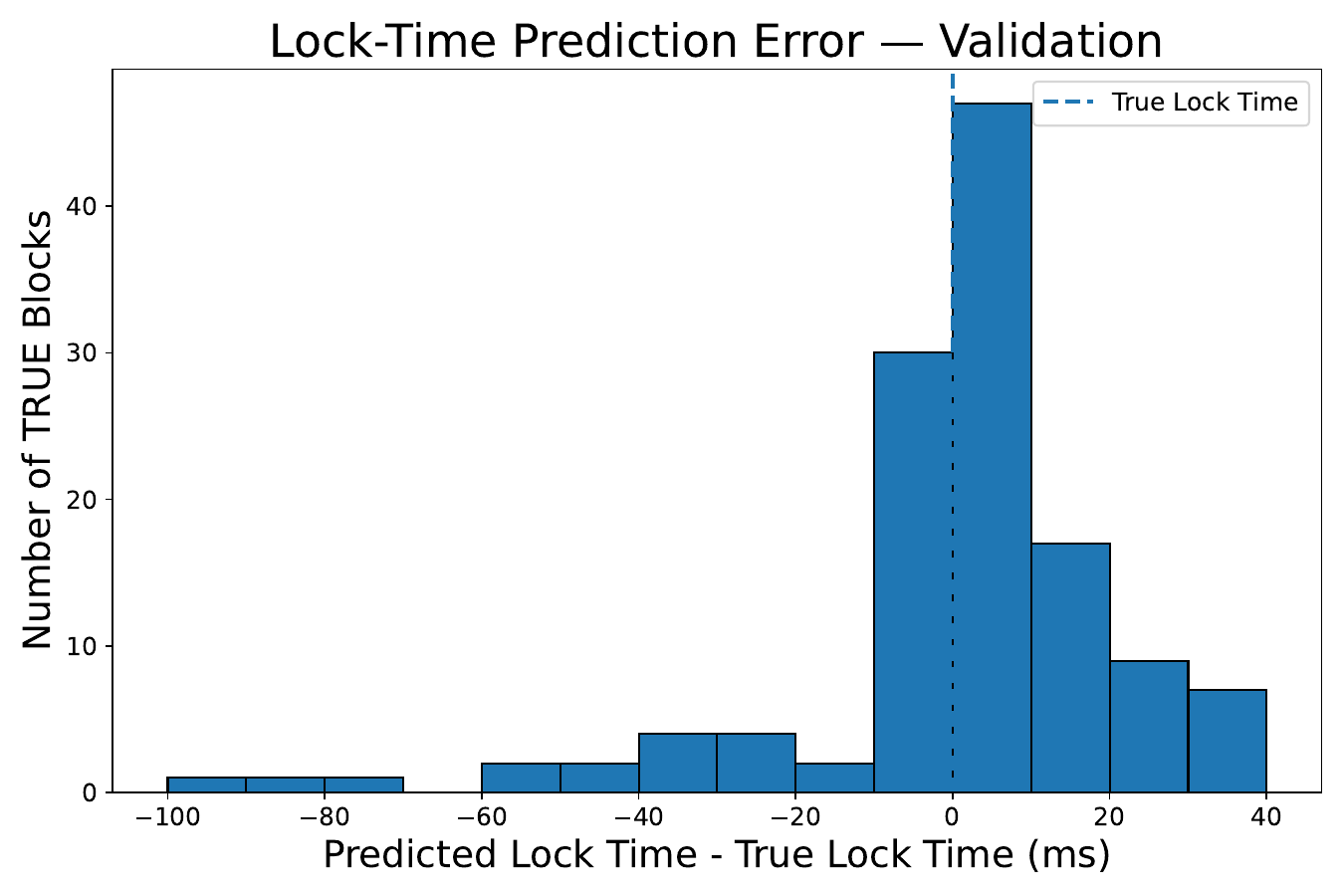}
\caption{This histogram displays the results of the LM time estimation algorithm applied to the true-positive test blocks after being tuned to the true-positive validation blocks. Outside of a few anomalous errors, the simple method closely times many of the LMs. All predictions terminated through the consecutive ridge-failure criterion rather than reaching the maximum search horizon of 100 ms beyond the end of the true-positive block.}
\label{Histogram_OFLMTimePRed}
\end{figure}

The Prominence, consecutive-failure limit, and shifting parameters were tuned on the validation set through a coarse parameter sweep, selecting for a balance of low error standard deviation and high accuracy within $\pm10$ ms. Figure \ref{Histogram_OFLMTimePRed} and Table \ref{TimeofBlockDetection} show the resulting performance.

\begin{table}[!ht]
\begin{center}
\begin{tabular}{||c c c||} 
 \hline
 Lock Time Estimation & Validation & Test \\ [0.5ex] 
 \hline\hline
  Correct Within $\pm10$ ms &  74.0\% & 69.1\%\\ 
 \hline
 Correct Within $\pm20$ ms &  82.7\% & 78.0\%\\ 
 \hline
 Correct Within $\pm30$ ms &  87.4\% & 82.9\%\\ 
 \hline
 Correct Within $\pm50$ ms &  96.1\% & 95.1\%\\ 
 \hline
 Correct Within $\pm100$ ms &  100\% & 99.2\%\\ 
 \hline
 Error Standard Deviation &  22.2 ms & 26.4 ms\\ 
 \hline
 Mean Absolute Error &  13.5 ms & 15.9 ms\\ 
 \hline
 Prominence &  0.1 & 0.1\\ 
 \hline
 Consecutive Failure Limit &  2 & 2\\ 
 \hline
 Shift &  +10 ms & +10 ms\\ 
 \hline
\end{tabular}
\end{center}
\caption{The analysis of using a simple ridge detection algorithm and following the ridge to termination has provided an accurate way to estimate the time of LM formation. This method was only applied to true-positive blocks of predicted Label 3s (using the blocks for each respective validation/test CNN). The parameters of the algorithm are listed along with the results; these parameters were tuned on the validation set and then applied unchanged to the test set.}
\label{TimeofBlockDetection}
\end{table}

%% file: Section5.tex
\section{Prediction Capability of the Model}\label{SectionPredictionCapability}
\begin{figure}[!ht]
\centering
\includegraphics[width=8.5cm,height=8.5cm,keepaspectratio]{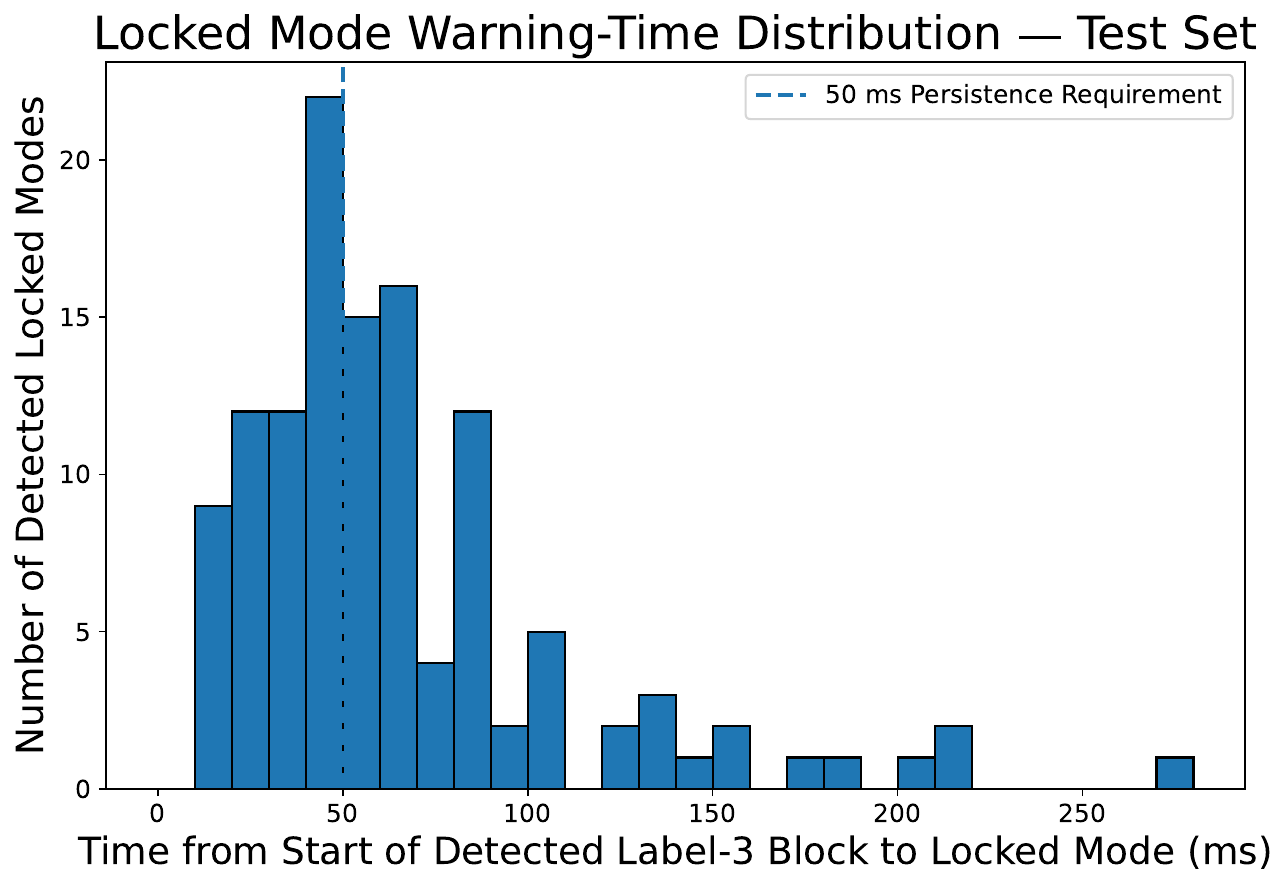}
\caption{This histogram outlines the spread of warning times (the locked mode time minus the time of the first predicted Label 3 in a true block) across the 121 test shots from their respective folds of the 10-fold cross-validation. A dashed vertical line is included at 50 ms to separate LMs for which the persistence requirement consumes the entire available warning time from those for which some advance warning remains.}
\label{Histogram_WarningTime}
\end{figure}

The model has been extensively trained and tuned to attain accurate detection levels; the prediction task, however, is more challenging for this approach. Table \ref{TableWarningTime} and Figure \ref{Histogram_WarningTime} outline how much warning time this model can give Plasma Control Systems (PCS) to respond to an imminent LM \cite{humphreys2005integrated}; unfortunately, it would appear that about half the LMs with prior rotation will lock before this method can raise the alert. In addition to this, the amount of warning time for the LMs that are detected early enough to alert the PCS may only amount to tens of ms. This problem arises due to the need to enforce a persistence filter of 50 ms to filter out false-positives.

This 50 ms restriction can be loosened for the sake of tokamak safety if a higher false-positive rate can be tolerated or if the procedure for preventing an LM is non-disruptive enough to warrant stronger reactions. Retuning of the CNN to provide longer lead times could improve this metric as well, but may come at the cost of a less effective LM detector. Alternatively, the Label 2 early-warning alert can be used to provide greater warning lead time. A similar analysis to what has already been performed with Label 3 could be performed with Label 2 in future work. The greatest limiting factor is built into the approach: the reactive observation of slowing MHD modes by RIP. This process can happen slowly or quickly, and it may not always be possible to confidently detect it in time using a spectrogram-based approach.

\begin{table}[!ht]
\begin{center}
\begin{tabular}{||c c c||} 
 \hline
 Warning Distribution & Validation & Test \\ [0.5ex] 
 \hline\hline
  0-50 ms &  49.6\% & 56.9\%\\ 
 \hline
 60-100 ms &  34.4\% & 31.7\%\\ 
 \hline
 110-200 ms &  15.2\% & 8.9\%\\ 
 \hline
 210-300 ms &  0.8\% & 2.4\%\\ 
 \hline
 300+ ms &  0\% & 0\%\\ 
 \hline
\end{tabular}
\end{center}
\caption{This table outlines the warning times from both the validation and test blocks that survived filtering. To ensure an emergency warning outputted by the model is accurate, the persistence filter dictates that it must last 50 ms. This delays the warning, resulting in about half of the LMs being predicted too late.}
\label{TableWarningTime}
\end{table}

%% file: Section6.tex
\section{Conclusion and Future Work}
This body of work has extended the existing overarching research goal of developing FPP-relevant tools for plasma state estimation. The capabilities and limitations of models trained on RIP data for LM detection and prediction have been thoroughly explored and benchmarked. The block-based detection technique has been shown to be quite accurate, identifying 88.5\% of LMs with a precision of 83.7\% and an F1 score of 86.0\%, while the lock time can be estimated with a mean absolute error of only 15.9 ms. While we were able to filter out many aberrations in the data, low-frequency, decelerating modes that do not ultimately lock remain difficult to distinguish from genuine locking events and may contribute to reduced precision. The prediction capability of LMs with RIP has proven to be a more difficult challenge than detection, limited by the required persistence of 50 ms of consistent emergency warning labels. Overall, the technique has thoroughly explored the capabilities of what can be done to analyze LMs without any magnetic diagnostics, established a viable method for LM detection and limited prediction, and benchmarked the results for future work. The technique is agnostic to the poloidal and toroidal mode number, only identifying the presence of any mode and analyzing its behavior.

Strict adherence to sound machine-learning principles was maintained through the use of a 10-fold cross-validation procedure designed to limit temporal and shot-based leakage between the training, validation, and test sets. In addition, validation data were used for model selection and hyperparameter tuning while preserving the held-out test data for unbiased evaluation. This approach provides a rigorous estimate of how the developed techniques generalize to unseen DIII-D discharges despite the limitations of the available dataset.

This work demonstrates that significant LM analysis can be performed using only a single interferometric chord measuring density fluctuations. These results are particularly promising for future FPP applications, where additional diagnostic measurements and larger training datasets may further improve model performance and predictive capabilities. Future work will expand on this to explore born-locked mode detection. Born-locked modes exhibit no prior rotation and therefore do not readily appear in the RIP spectrograms used by this approach. Without magnetic diagnostics, addressing born-locked modes may require spatially distributed measurements capable of observing emergent mode structures in the plasma; the electron cyclotron emission imaging (ECEI) diagnostic at DIII-D is a strong candidate for investigating this capability. Greater predictive power may also come from ECEI or another FPP-relevant diagnostic capable of providing additional information about the conditions leading to LM formation. The Label 2 early-warning class can also be explored further, using either RIP or additional diagnostics such as ECEI, to provide greater warning times and potentially more reliable predictions.

%% file: DataAvailibility.tex
\section{Data Availability Statement}
All data that support the findings of this study are included within the article (and any supplementary files). Non-DIII-D data and labels can be provided upon request.

%% file: Section7.tex
\section{Acknowledgment}
This material is based upon work supported by the U.S. Department of Energy, Office of Science, Office of Fusion Energy Sciences, using the DIII-D National Fusion Facility, a DOE Office of Science user facility, under Awards DE-FC02-04ER54698, DE-FG02-05ER54809, DE-FG02-97ER54415, DE-SC0019352, DE-SC0021150, and Next Step Fusion S.a.r.l. The authors would like to thank Justin Burzachiello, Jayson Barr, Nathan Richner, and Tyler Cote for fruitful discussions.

%% file: Section8.tex
\section*{Disclaimer}
This report was prepared as an account of work sponsored by an agency of the United States Government. Neither the United States Government nor any agency thereof, nor any of their employees, makes any warranty, express or implied, or assumes any legal liability or responsibility for the accuracy, completeness, or usefulness of any information, apparatus, product, or process disclosed, or represents that its use would not infringe privately owned rights. Reference herein to any specific commercial product, process, or service by trade name, trademark, manufacturer, or otherwise does not necessarily constitute or imply its endorsement, recommendation, or favoring by the United States Government or any agency thereof. The views and opinions of authors expressed herein do not necessarily state or reflect those of the United States Government or any agency thereof.